\documentstyle[sprocl]{article}

\def\Journal#1#2#3#4{{#1} {\bf #2}, #3 (#4)}

\def\PLA{{\em Phys. Lett.}  A}

\def\be{\begin{equation}}
\def\ee{\end{equation}}
\def\bea{\begin{eqnarray}}
\def\eea{\end{eqnarray}}

\begin{document}


\title{ON $qp$-DEFORMATIONS IN STATISTICAL MECHANICS OF BOSONS IN $D$ 
DIMENSIONS} 

\author{ M.R. KIBLER and J. MEYER }

\address{Institut de Physique Nucl\'eaire de Lyon,\\
IN2P3-CNRS et Universit\'e Claude Bernard,\\
43, Boulevard du 11 Novembre 1918,\\
F-69622 Villeurbanne Cedex, France}

\author{ M. DAOUD }

\address{Laboratoire de Physique Th\'eorique,\\
Facult\'e des Sciences, Universit\'e Mohammed V,\\
Avenue Ibn Batouta,\\
B.P.~1014, Rabat, Morocco}


\maketitle\abstracts{
The Bose distribution for a gas of nonrelativistic free bosons is
derived in the framework of $qp$-deformed second quantization.
Some thermodynamical functions for such a system in $D$ dimensions 
are derived. Bose-Einstein condensation is discussed in terms of the 
parameters $q$ and $p$ as well as a parameter $\nu_0^{\prime}$ which
characterizes the representation space of the oscillator algebra.}

\vskip 6.5 true cm 

\noindent
This paper is based on a lecture (given by M.R. Kibler) to the IVth 
International
School on Theoretical Physics ``Symmetry and Structural Properties of Condensed
Matter'' (Zaj\c aczkowo, Poland, 29 August - 4 September 1996). 
It will be published in {\em Symmetry and Structural Properties of
Condensed Matter}, eds. T. Lulek, B. Lulek and W. Florek (World Scientific,
Singapore, 1997). 

\vfill\eject

\title{ON $qp$-DEFORMATIONS IN STATISTICAL MECHANICS OF BOSONS IN $D$ 
DIMENSIONS} 

\author{ M.R. KIBLER, J. MEYER }

\address{Institut de Physique Nucl\'eaire de Lyon,\\
IN2P3-CNRS et Universit\'e Claude Bernard,\\
43, Boulevard du 11 Novembre 1918,\\
F-69622 Villeurbanne Cedex, France}

\author{ M. DAOUD }

\address{Laboratoire de Physique Th\'eorique,\\
Facult\'e des Sciences, Universit\'e Mohammed V,\\
Avenue Ibn Batouta,\\
B.P.~1014, Rabat, Morocco}


\maketitle\abstracts{
The Bose distribution for a gas of nonrelativistic free bosons is
derived in the framework of $qp$-deformed second quantization.
Some thermodynamical functions for such a system in $D$ dimensions 
are derived. Bose-Einstein condensation is discussed in terms of the 
parameters $q$ and $p$ as well as a parameter $\nu_0^{\prime}$ which
characterizes the representation space of the oscillator algebra.}

\section{Introduction}

The use of quantum groups and quantum algebras is now largely displayed in
theoretical physics. From an elementary point of view, quantum algebras may be
thought of as 
$q$- (or $qp$)-deformations of Lie algebras. Roughly speaking, there
are two types of applications of $q$-deformations to Physics (cf.~Ref.~1). 
Applications of type A dot not rely on phenomenology. For example, applications
of type A concern 
the quantum inverse scattering method, the quantum Yang-Baxter equation
and a true definition of space-time. On the other hand, applications of type B
are entirely phenomenological. Along this vein, we may mention the use of the
two-parameter quantum algebra $ U_{qp} ( {\rm u}_2) $ (a Hopf algebra) 
to rotational  collective  dynamics 
of atomic nuclei.$^2$ For applications of type B, we immediately face
the problems of the physical interpretation and the nonuniversality of the
deformation parameter(s). 

The considerations above can be applied to deformed oscillator algebras. 
Indeed, oscillator algebras and quantum algebras may be connected since the 
$q$- (or $qp$)-deformed creation and annihilation operators, which are defined
in an oscillator algebra, may serve for constructing realizations of 
$q$- (or $qp$)-deformed Lie algebras. 

It is the aim of this lecture to examine the interest of using $qp$-deformed
creation and annihilation operators in statistical mechanics of bosons in
arbitrary dimension. In other words, we ask the following question: What are the
implications, at the level of the Bose distribution  
and the Bose-Einstein condensation, 
of replacing ordinary boson operators by $qp$-deformed creation
and annihilation operators? Furthermore, is it possible to test 
a deformed Bose distribution and to find 
a physical interpretation of the deformation parameter(s)?

This subject was already investigated, mainly in the case of 
$q $-deformed boson operators, by many authors.$^{3-18}$ The case of 
$qp$-deformed boson operators was considered in 
Refs.~17 and 18. Most of the works in Refs.~3 to 18 use the Fock
representation$^{19-25}$ for the deformed oscillator algebra 
inherent to the considered 
creation and annihilation operators. In contradistinction, the representation
introduced by  Rideau$^{26}$  (see also Ref.~1) for the $q$-deformed
oscillator algebra is used in Ref.~16. 

We shall deal in this paper with both the Fock and the Rideau representations
of deformed oscillator algebras. These representations are 
described in Sec.~2. Then, Sec.~3 is devoted to $qp$-deformations of the Bose
distribution and its consequence for the Bose-Einstein condensation 
phenomenon. Finally, some conclusions are presented in Sec.~4. Two appendices
close this paper. In particular, Appendix B concerns a $qp$-deformed
correlation factor for the radiation field. 

This work takes its origin in Refs.~9 and 18. It is however to be noticed 
that some of the developments and the conclusions given here 
substancially differ from the ones in Refs.~9 and 18. 

\section{Deformed Oscillator Algebra}

Following Refs.~1 and 26, we characterize the algebra $W_{qpQ}$ by 
three operators $a$, $a^+$ and $N$ such that 

{$\quad \bullet$} the operator $a^+$ is the adjoint of $a$,

{$\quad \bullet$} the operator $N$ is self-adjoint with a discrete 
nondegenerate spectrum, and 

{$\quad \bullet$} the operators $a$, $a^+$ and $N$ satisfy the commutation
relations 
\begin{equation}
[N,a  ] = -a,   \quad 
[N,a^+] =  a^+, \quad 
aa^+ - Qa^+a = {1 \over {q-p}} [q^N (q-Q) - p^N (p-Q)]
\label{eq:aa+N}
\end{equation}
where the three parameters $q$, $p$ and $Q$ are fixed parameters 
taken (a priori) in the field of complex numbers. 

The algebra $W_{qpQ}$ spanned by the deformed {\em annihilation} ($a$), 
{\em creation} ($a^+$) and {\em number} ($N$) operators is called a deformed
oscillator algebra. In the special cases $Q=q$ and $Q=p$, 
the operators $a$ and $a^+$ are referred to as $qp$-boson
operators. They entail the particular cases of the $q$-boson operators as used
in mathematics (when $p=1$) and in physics (when $p=q^{-1}$). 
Four particular cases are of interest for the applications 
(see also Refs.~19 to 26). 

1) The case $Q=q$, $p=1$: This case corresponds to
\begin{equation}
[N,a  ] = -a,   \qquad 
[N,a^+] =  a^+, \qquad 
aa^+ - qa^+a = 1 
\label{eq:bm}
\end{equation}
i.e. to $q$-bosons as used in the mathematical literature 
(see also Refs.~19 to 21 and 25). 

2) The case $Q=q$, $p=q^{-1}$: This case corresponds to
\begin{equation}
[N,a  ] = -a,   \qquad 
[N,a^+] =  a^+, \qquad 
aa^+ - qa^+a = q^{-N} 
\label{eq:bp1}
\end{equation}
i.e. to $q$-bosons as used in the physical literature.$^{22,23}$

3) The case $Q=q^{-1}$, $p=q^{-1}$: This case corresponds to
\begin{equation}
[N,a  ] = -a,   \qquad 
[N,a^+] =  a^+, \qquad 
aa^+ - q^{-1}a^+a = q^{N} 
\label{eq:bp2}
\end{equation}
i.e. to $q$-bosons as used in the physical literature.$^{22,23}$

4) The case $Q=0$: This case corresponds to
\begin{equation}
[N,a  ] = -a,   \qquad 
[N,a^+] =  a^+, \qquad 
aa^+ = { { q^{N+1} - p^{N+1} } \over {q-p} } 
\label{eq:MB}
\end{equation}
which admits two interesting subcases corresponding 
to either $p=1$ or $p=q^{-1}$. 

The nondeformed oscillator algebra corresponds to 
($Q=q$,      $p=1     $, $q \to 1$) or
($Q=q$,      $p=q^{-1}$, $q \to 1$) or
($Q=q^{-1}$, $p=q^{-1}$, $q \to 1$). Note that the cases 
($Q=0$,      $p=1     $, $q \to 1$) and 
($Q=0$,      $p=q^{-1}$, $q \to 1$) correspond to the exotic limiting case
where $aa^+ = 1$. 

The Hilbertean irreducible representations of the oscillator algebra 
$W_q \equiv W_{qq^{-1}q}$ were dealt in details by Rideau.$^{26}$ 
The Hilbertean irreducible representations of the more general algebra 
$W_{qpQ}$ can be derived equally well. It is enough to say here that the 
relations
\begin{eqnarray}
a   \> |n \rangle &=& \left( Q^{n}   \lambda_0 
+ p^{\nu_0} {{Q^{n}   - p^{n}}   \over {q-p}}
+ q^{\nu_0} {{q^{n}   - Q^{n}}   \over {q-p}} \right)^{1 \over 2}
\; |n - 1 \rangle \nonumber\\
a^+ \> |n \rangle &=& \left( Q^{n+1} \lambda_0 
+ p^{\nu_0} {{Q^{n+1} - p^{n+1}} \over {q-p}}
+ q^{\nu_0} {{q^{n+1} - Q^{n+1}} \over {q-p}} \right)^{1 \over 2}
\; |n + 1 \rangle \nonumber\\ 
N   \> |n \rangle &=& (\nu_0 + n) \; |n \rangle 
\label{eq:rep} 
\end{eqnarray} 
provide us with a {\it formal} representation of $W_{qpQ}$. 
We thus obtain an infinity of Hilbertean irreducible representations 
of $W_{qpQ}$ that may be arranged into two types:
$R(\nu_0^{\prime}) $ with $n \in {\bf N}$ and 
$S(\lambda_0,\nu_0)$ with $n \in {\bf Z}$. We shall deal in this work only with 
   the (Fock) representation $R(0)             $ of the algebra $W_{qpq     }$ 
and 
   the (Fock) representation $R(\nu_0^{\prime})$, with $\nu_0^{\prime} \ne 0$, 
                                                 of the algebra $W_q$. 
We shall not use the case of 
the (nonFock) representation $S(\lambda_0,\nu_0)$. This representation which
has no limit at $q=p=Q=1$ is briefly discussed in Appendix A.  

The (deformed) Fock 
representation  $R(0)$  of $W_{qpq}$ 
is obtained from Eq.~(\ref{eq:rep}) by taking 
$\lambda_0 = \nu_0 = 0$ and $n \in {\bf N}$. Then, we have 
\begin{eqnarray} 
a   \; |n \rangle &=& {\sqrt {[[n]]_{qp}    }} \; |n - 1 \rangle,
\nonumber\\ 
a^+ \; |n \rangle &=& {\sqrt {[[n + 1]]_{qp}}} \; |n + 1 \rangle, 
\nonumber\\
N   \; |n \rangle &=& n                        \; |n     \rangle 
\label{eq:Fock} 
\end{eqnarray} 
with $a \; | 0 \rangle \; = \; 0$. In Eq.~(\ref{eq:Fock}), we use the notation
\begin{equation}
[[x]]_{ qp } \; = \; { {q^x - p^x} \over {q - p} }
\label{eq:qp-nbre} 
\end{equation}
It is to be observed that the operator $N$ is a nondeformed 
operator that coincides with the usual number 
operator (in the sense that its spectrum is $\bf N$). Note that 
\begin{equation}
[[x]]_{ q, p=     1 } \; = \; { {q^x -      1} \over {q -      1} }, \qquad
[[x]]_{ q, p=q^{-1} } \; = \; { {q^x - q^{-x}} \over {q - q^{-1}} }
\label{eq:q(mp)-nbre} 
\end{equation}
The case of the nondeformed (i.e.~usual) Fock representation is reached for 
($p=     1$, $q \to 1$) or
($p=q^{-1}$, $q \to 1$). Finally, the case ($p = 1$, $q = 0$) corresponds to
the exotic situation where 
\begin{equation} 
a   \; |n \rangle \; =   \;      |n - 1 \rangle, \quad
a^+ \; |n \rangle \; =   \;      |n + 1 \rangle, \quad
N   \; |n \rangle \; =   \; n \; |n     \rangle 
\label{eq:Marek} 
\end{equation} 
Going back to the representation $R(0)$ (given by Eq.~(\ref{eq:Fock})) of 
$W_{qpq}$, we observe that 
\begin{equation} 
a^+  a = [[N    ]]_{qp}, \qquad  
a a^+  = [[N + 1]]_{qp} 
\label{eq:NN+1} 
\end{equation} 
so that the relation 
\begin{equation}
aa^+ - pa^+a = q^{N} 
\label{eq:pqN} 
\end{equation} 
holds for this representation in addition to the defining relation 
\begin{equation}
aa^+ - qa^+a = p^{N} 
\label{eq:qpN} 
\end{equation} 
Equations (\ref{eq:pqN}) and (\ref{eq:qpN}) clearly show 
that the representation $R(0)$ of 
$W_{qpq}$ exhibits the $q \leftrightarrow p$ symmetry.  

The (deformed) Fock representation $R(\nu_0^{\prime})$ of $W_q$ 
is obtained from Eq.~(\ref{eq:rep}) by taking 
$\lambda_0 = 0$, $\nu_0 = \nu_0^{\prime} \ne 0$ and $n \in {\bf N}$. 
Then, we have
\begin{eqnarray}
a   \> |n \rangle &=& q^{-{\nu_0^{\prime} \over 2}} {\sqrt{[n  ]_q}}
                         \; |n - 1 \rangle \nonumber\\
a^+ \> |n \rangle &=& q^{-{\nu_0^{\prime} \over 2}} {\sqrt{[n+1]_q}}
                         \; |n + 1 \rangle \nonumber\\
N   \> |n \rangle &=& (\nu_0^{\prime} + n) 
                         \; |n     \rangle 
\label{eq:RidFock} 
\end{eqnarray} 
with $a \; | 0 \rangle \; = \; 0$. In Eq.~(\ref{eq:RidFock}), 
we use the notation
\begin{equation}
[x]_q \; = \; [[x]]_{ q, p=q^{-1} } 
\label{eq:q-nbre-bis} 
\end{equation}
and we have $q > 1$ and $\nu_0^{\prime} \in {\bf R}$. 
It is to be observed that the operator $N$ does not 
coincide with the usual number operator; the usual number operator is now 
$N - \nu_0^{\prime}$ (in the sense that its spectrum is $\bf N$). As an
important result, we have that if $q \to 1^+$ then $\nu_0^{\prime} \to 0$; in
other words, we can write  
\begin{equation}
\lim_{q \to 1^+} R(\nu_0^{\prime}) = {\rm usual \ Fock \ representation} 
\label{eq:limite} 
\end{equation} 
For the representation $R(\nu_0^{\prime})$, we have 
\begin{equation} 
a^+  a    = q^{-{\nu_0^{\prime}}} [N     - \nu_0^{\prime}]_{q}, \qquad  
a    a^+  = q^{-{\nu_0^{\prime}}} [N + 1 - \nu_0^{\prime}]_{q} 
\label{eq:NN+1bis} 
\end{equation} 
Consequently, we verify that the relations 
\begin{equation}
aa^+ - qa^+a = q^{-N}, \qquad aa^+ - q^{-1}a^+a = q^{N - 2 \nu_0^{\prime}} 
\label{eq:qpNpqN} 
\end{equation} 
hold in $W_q$. 
Equation (\ref{eq:qpNpqN}) indicates
that the representation $R(\nu_0^{\prime})$ of $W_q$ does not present 
the $q \leftrightarrow q^{-1}$ symmetry.

\section{Bose Distribution and Bose-Einstein Condensation}

 Let us consider a gas of nonrelativistic free bosons. Its Hamiltonian 
 (in a second-quantized form) reads
 \begin{equation}
 H := \sum_k H_k, \qquad H_k := (E_k - \mu) \nu_k 
 \label{eq:(3)}
 \end{equation}
 In Eq.~(\ref{eq:(3)}), 
 $\mu$ is the chemical potential while $E_k$ and $\nu_k$ are the kinetic 
 energy of a boson and the number operator for the bosons, 
 in the $k$ mode, respectively. The Bose factor for the $k$ mode is then
 \begin{equation}
 f_k := {1\over Z} {\rm tr}  
                                   \big ( {\rm e}^{-\beta H} a^+_k a_k \big ) 
 \label{eq:(4)}
 \end{equation}
 where 
 \begin{equation}
 Z := {\rm tr} \left( {\rm e}^{-\beta H} \right)
 \label{eq:Z}
 \end{equation}
 is the partition function and $\beta = (k_B T)^{-1}$ 
 the reciprocal temperature. 

 It is essential to note that Eqs.(\ref{eq:(3)})-(\ref{eq:Z}) 
 have the same form as in the nondeformed case. 
 They correspond to the basic 
 formulas for the statistical mechanics of bosons. Here, we do
 not deform the basic formulas. The trace and the exponential 
 in (\ref{eq:(4)}) and 
    (\ref{eq:Z}) are nondeformed functions. 
 The deformation is introduced via the use of annihilation 
 ($a_k$) and creation $(a^+_k)$ 
 operators that satisfy $qp$-deformed commutation relations. The
 form chosen for $H_k$ in Eq.~(\ref{eq:(3)}) is nothing but the common 
 one where the usual number operator $\nu_k$ 
 takes its eigenvalues in ${\bf N}$. The
 deformation cannot enter the theory at this level by replacing 
 $\nu_k$ by $a_k^+a_k$ in (\ref{eq:(3)}) since the operator 
 $a_k^+a_k$ has not its eigenvalues 
 in ${\bf N}$. On the contrary, we must keep $a_k^+a_k$ rather than
 $\nu_k$ in (\ref{eq:(4)}). Indeed, should we use 
 $\nu_k$ in place of $a_k^+a_k$ in
 (\ref{eq:(4)}), we would get a nondeformed value of the Bose factor. 
 As a conclusion, Eqs.~(\ref{eq:(3)})-(\ref{eq:Z}) 
 constitute a reasonable starting point for a (minimal)
 deformation of the Bose distribution.  

 We are now in a position to derive several types of deformed Bose
 distributions. The deformation is explicitly introduced by using either 
 the representation $R(0)$ of $W_{qpq}$ (see Eq.~(\ref{eq:Fock})) 
 or the representation $R(\nu_0^{\prime})$, with $\nu_0^{\prime} \ne 0$, 
 of $W_q$ (see Eq.~(\ref{eq:RidFock})). The two representations can
 be treated (in a formal way) on the same footing by considering 
 (cf.~Eqs.~(\ref{eq:Fock}) and (\ref{eq:RidFock})) the representation 
 \begin{eqnarray}
 a_k   \> |n \rangle &=& p^{{\nu_0^{\prime} \over 2}} {\sqrt{[[n  ]]_{qp}}}
                            \; |n - 1 \rangle \nonumber\\
 a_k^+ \> |n \rangle &=& p^{{\nu_0^{\prime} \over 2}} {\sqrt{[[n+1]]_{qp}}}
                            \; |n + 1 \rangle \nonumber\\
 N_k   \> |n \rangle &=& (\nu_0^{\prime} + n) 
                            \; |n     \rangle 
 \label{eq:unif} 
 \end{eqnarray} 
 of the algebra $W_{qpq}$. We assume (mathematical hyphothesis): 
 $\left\{ a_k, a_k^+, N_k \right\}$ and 
 $\left\{ a_j, a_j^+, N_j \right\}$ are two commuting sets when $j \ne k$. 
 Furthermore, we take (physical hyphothesis):
 \begin{equation}
 \nu_k := N_k - \nu_0^{\prime}
 \label{eq:nuN}
 \end{equation}
 As a trivial result, we have
 \begin{equation}
 Z = \prod_{k} { 1 \over 1 - {\rm e}^{-\eta} } 
 \label{eq:(5)}
 \end{equation}
 where
 \begin{equation}
 \eta := \beta (E_k - \mu)
 \label{eq:(5bis)}
 \end{equation}
 The partition function $Z$ is thus independent of the
 deformation parameters $q$ and $p$. 
 Then, the Bose factor $f_k$ can be written 
 \begin{equation} 
 f_k = p^{ \nu_0^{\prime} } 
 (1 - {\rm e}^{-\eta}) {\rm tr}  
                               \big ( {\rm e}^{-\eta (N_k - \nu_0^{\prime})} 
 [[N_k - \nu_0^{\prime}]]_{qp} \big ) 
 \label{eq:(6a)} 
 \end{equation}
 which can be shown to converge in each of the following five cases:

 {$\quad \bullet$} $(q,p) \in {\bf R}^+ \times {\bf R}^+$ with
 $0 < q < {\rm e}^{-\beta \mu}$ and $0 < p < {\rm e}^{-\beta \mu}$

 {$\quad \bullet$} $(q,p) \in {\bf R}^+ \times {\bf R}^+$ with
 $p=q^{-1}$ and ${\rm e}^{ \beta \mu} < q < {\rm e}^{-\beta \mu}$

 {$\quad \bullet$} $(q,p) \in {\bf R}^+ \times {\bf R}^+$ with
 $p=1$ and $0 < q < {\rm e}^{-\beta \mu}$

 {$\quad \bullet$} $(q,p) \in {\bf C} \times {\bf C}$ with
 $p=\bar q$ and $0 < |q| < {\rm e}^{-\beta \mu}$

 {$\quad \bullet$} $(q,p) \in S^1 \times S^1$ with
 $p=\bar q = q^{-1}$.

 The latter two cases, although acceptable from the mathematical point of view,
 do not lead to physically acceptable statistical distributions. 

 As a formal result, Eq.~(\ref{eq:(6a)}) yields 
 \begin{equation} 
f_k = p^{ \nu_0^{\prime} } \bigg (
{q-1\over q-p} {1\over {\rm e}^\eta - q} +
{p-1\over p-q} {1\over {\rm e}^\eta - p} \bigg )
 \label{eq:(9a)}
 \end{equation}
or alternatively 
 \begin{equation}
f_k = p^{ \nu_0^{\prime} } { {\rm e}^{\eta} - 1 \over 
              ({\rm e}^{\eta} - q) 
              ({\rm e}^{\eta} - p) } 
 \label{eq:(10)}
 \end{equation}
In the limiting situation where $q \to 1$ and $p \to 1$, 
we recover the classical Bose distribution
\begin{equation} 
\Phi_k = \frac{1}{ {\rm e}^\eta -1 }
\label{eq:BOSE}
\end{equation}
The distribution $f_k$ can be rewritten as
 \begin{equation}
f_k = p^{ \nu_0^{\prime} } \bigg [
             {q-1\over q-p} {\Phi_k \over 1 + (1-q)\Phi_k } 
           + {p-1\over p-q} {\Phi_k \over 1 + (1-p)\Phi_k } \bigg ]
 \label{eq:(9b)}
 \end{equation}
in term of $\Phi_k$.

For the purpose of practical applications, it is useful 
to know the development of the distribution $f_k$ in integer 
series. In this respect, we have 
 \begin{equation}
f_k = p^{ \nu_0^{\prime} }
\sum^\infty_{j=0} {\rm e}^{ - \eta (j+1) } \left ( {q-1\over q-p} q^j + 
                                                   {p-1\over p-q} p^j \right ) 
 \label{eq:(6b)}
 \end{equation}
which reduces to the expansion $\sum^{\infty}_{j=1} {\rm e}^{ -\eta j }$ of the
ordinary Bose factor when $q \to 1$ and $p \to 1$. 

 In the thermodynamical limit, the energy spectrum of the system of 
bosons may be considered as a continuum. Thus, $f_k$ is 
replaced by
the $qp$-dependent factor $f(\epsilon)$ which is $f_k$ 
with $E_k = \varepsilon$. Therefore, in physical applications, 
we have to consider integrals of the type
 \begin{equation}
J_s := \int^\infty_0 \varepsilon^s f(\varepsilon) d\varepsilon 
 \label{eq:(11)}
 \end{equation}
By using the development (\ref{eq:(6b)}), Eq.~(\ref{eq:(11)}) leads to 
 \begin{equation}
J_s = \Gamma (s+1) (k_B T)^{s+1} \sigma (s+1)_{qp}, \qquad s>-1 
 \label{eq:(12)}
 \end{equation}
where
 \begin{equation}
\sigma(s+1)_{qp} = p^{ \nu_0^{\prime} } \sum^\infty_{j=0} 
{{\rm e}^{ \beta \mu (j+1)} \over (j+1)^{s+1}} 
\big ( [[j+1]]_{qp} - [[j]]_{qp} \big ) 
 \label{eq:(13)}
 \end{equation}
In Eq.~(\ref{eq:(12)}), $\Gamma$ is the Euler integral of the 
second type. In the case where $\mu = 0$, we have
$\sigma (s+1)_{qp} \rightarrow \zeta (s+1)$ for $q \to 1$ and $p \to 1$, 
where $\zeta(s+1)$ is the Riemann Zeta series (that converges for $s>0$).

We now derive some thermodynamical quantities, in 
$D$ dimensions, for a free gas of $N(D)$ bosons contained in 
a volume $V(D)$. The density 
\begin{equation}
\rho(D) := \frac{N(D)}{V(D)}
\label{eq:dens}
\end{equation}
is given by$^{9}$ 
 \begin{equation}
\rho (D) 
= N_0(D) \Gamma ({D \over 2}) (k_B T)^{{D \over 2}} \sigma ({D \over 2})_{qp} 
 \label{eq:(14)}
 \end{equation}
where
 \begin{equation}
N_0(D) := 
{ 1 \over 2 (2 \pi)^{D \over 2} }  
{ D \over {\Gamma({D \over 2} + 1)} } 
g {m^{D \over 2} \over \hbar^D} 
 \label{eq:(15)}
 \end{equation}
In Eq.~(\ref{eq:(15)}), 
$g$ is the degree of (spin) degeneracy and $m$ the mass of a
boson. The total energy can then be calculated to be
 \begin{equation}
E(D) = N_0(D) V(D)               \Gamma \big ({D \over 2} + 1 \big )
       (k_B T)^{{D \over 2} + 1} \sigma \big ({D \over 2} + 1 \big )_{qp} 
 \label{eq:(16)}
 \end{equation}
The specific heat at constant volume easily follows from
$C_V = \big ( {\partial E \over \partial T } \big )_V$. We obtain
 \begin{equation}
C_V(D) = {D \over 2} N(D) k_B 
\bigg [ 
       \big  ( {D \over 2} + 1 \big ) 
{\sigma \big ( {D \over 2} + 1  \big )_{qp} \over
 \sigma \big ( {D \over 2}      \big )_{qp}}
             - {D \over 2} 
{\sigma \big ( {D \over 2}      \big )_{qp} \over
 \sigma \big ( {D \over 2} - 1  \big )_{qp}}
\bigg ] 
 \label{eq:(17)}
 \end{equation}
Furthermore, the entropy is 
 \begin{equation}
S(D) = N(D) k_B 
\bigg [ 
        \big ( {D \over 2} + 1 \big ) 
{\sigma \big ( {D \over 2} + 1 \big )_{qp} \over
 \sigma \big ( {D \over 2}     \big )_{qp}} - {\mu \over k_B T} 
\bigg ]
 \label{eq:(18)}
 \end{equation}
Note that the state equation for the gas of bosons is
 \begin{equation}
p(D) = {2 \over D} { E(D) \over V(D) } 
 \label{eq:(19)}
 \end{equation}
so that the pressure $p(D)$ assumes the same form as in the nondeformed case. 

Finally, other thermodynamical quantities may be determined 
in a simple manner  by  using  the  thermodynamic potential 
 \begin{equation}
\Omega(D) = - {2 \over D} N_0(D) V(D) J_{{D \over 2}}
          = - {2 \over D} E(D) 
 \label{eq:(20)}
 \end{equation}
where $J_{{D\over 2}}$ is given by (\ref{eq:(12)}) and (\ref{eq:(13)}).

We now examine the condensation of a system 
   of bosons in $D$ dimensions. 
The corresponding density for such a system 
is  given  by  (\ref{eq:(14)}) which can be
rewritten as 
 \begin{equation}
\rho (D) = N_0 (D) J_{ {D \over 2} - 1 }
 \label{eq:(21)}
 \end{equation}
as a function of the integral $J_{ {D \over 2} - 1 }$. 

As in the classical case (i.e.~$q=p=1$), we have to define a 
procedure for generating the critical Bose temperature $T_c(D)$. 
For this purpose, we follow Ref.~9 by taking
\begin{equation}
\beta \mu = \beta \mu_1 - \ln q, \qquad q>1 
\label{eq:mu-q}
\end{equation}
where $\mu_1$ is a negative constant corresponding to the chemical potential in
the classical case.This $q$-dependence of the chemical potential $\mu$ makes it
possible to have a good behaviour of the distribution $f_k$ when 
$\mu_1 \to 0^-$ only for the two situations $p=q^{-1}$ and $p=1$. 
Therefore, in the remaining part of this section, the parameter $p$ stands for
                                     either $p=q^{-1}$ or  $p=1$. The
Bose-Einstein condensation is then obtained by introducing $\mu_1 = 0$ in 
Eq.~(\ref{eq:(21)}). Thus, we find that the critical temperature 
below which we obtain Bose-Einstein condensation is given by 
\begin{equation}
T_c(D)_{qp} = p^{ - \frac{2}{D} \nu_0^{\prime} } {1 \over k_B} 
\bigg [ {\rho(D) \over N_0(D)} 
{1 \over \Gamma ({D \over 2}) \sigma_0 ({D \over 2})_{qp}} 
\bigg ]^{2 \over D} 
\label{eq:(22)}
\end{equation}
where
\begin{equation}
\sigma_0({D \over 2})_{qp} := \frac{ 1 } { q+1 } \sum_{j=0}^\infty  
{1 + q^{-2j-1} \over (j+1)^{D \over 2}} 
\quad {\rm for} \quad p = q^{-1} 
\quad {\rm and} \quad 1 < q < {\rm e}^{- \beta \mu }
\label{eq:(23-1)}
\end{equation}
or
\begin{equation}
\sigma_0({D \over 2})_{qp} := \frac{ 1 } { q } \sum_{j=0}^\infty  
{1 \over (j+1)^{D \over 2}} 
\quad {\rm for} \quad p = 1 
\quad {\rm and} \quad 1 < q < {\rm e}^{- \beta \mu }
\label{eq:(23-2)}
\end{equation}
We thus end up with three models: a two-parameter model 
M$_1$ with ($\nu_0^{\prime} 
                         \in {\bf R}^*$, $p=q^{-1}$, $1<q<{\rm e}^{-\beta\mu}$)
and two one-parameter models, viz. 
M$_2$ with ($\nu_0^{\prime} 
                          =          0$, $p=q^{-1}$, $1<q<{\rm e}^{-\beta\mu}$)
and 
M$_3$ with ($\nu_0^{\prime} 
                          =          0$, $p=     1$, $1<q<{\rm e}^{-\beta\mu}$).
For each of these models, the Bose-Einstein condensation phenomenon takes place
only when the dimension $D$ is greater than 2, exactly as in the classical
case. (The series in Eqs.~(\ref{eq:(23-1)}) and 
                          (\ref{eq:(23-2)}) converge for $D \ge 3$.) 
The models M$_1$, M$_2$ and M$_3$ 
can be easily tested for $D=3$ by comparing for each model the theoretical and
experimental temperatures $T_c(3)$ for $^4$He superfluid in phase II. For the
models M$_2$ and M$_3$, we obtain 
\begin{equation}
T_c(3)_{qp} > T_c(3)_{\rm classical} > T_c(3)_{\rm exp} \sim 2.17~{\rm K} 
\label{eq:ine}
\end{equation}
Therefore, the models M$_2$ and M$_3$ give results which are worse than the one
given by the classical model (corresponding to $q=p=1$) for which we have
\begin{equation}
T_c(3)_{\rm classical} \equiv T_c(3)_{q=1, p=1} = {2 \pi \hbar^2 \over m k_B} 
\left [ {\rho(3) \over 2.612 g} \right ]^ {2 \over 3}
\label{eq:(24)}
\end{equation}
Unlike the models  M$_2$  and  M$_3$, for the model M$_1$, 
it is possible to find couples ($\nu_0^{\prime},q$) or which
$T_c(3)_{q, p=q^{-1}}$ is in agreement with the experimental value. As
a point of fact, our test rules out the models M$_2$ and M$_3$. 

\section{Conclusions}

Among the various $qp$-deformations of the Bose factor studied in this work,
only the ones corresponding to $p=q^{-1}$ or $p=1$ yield acceptable
statistical distributions for bosons. However, there is no $q$-deformation,
with $\nu_0^{\prime} = 0$, of the Bose factor that leads to a correct value of
the Bose-Einstein transition temperature. This has been tested for $^4$He but
we note that $T_c(3)_{qp} > T_c(3)_{\rm classical}$ holds in general. The
latter result generally applies to other $q$-deformations, available in the
literature,$^{4,5,10,15,16}$ based on the use of
the representation $R(0)$ of the oscillator algebra $W_q$. 

The model M$_1$ deserves more optimistic concluding remarks. As a matter of
fact, for this model, corresponding to 
($\nu_0^{\prime} < 0$, $p=q^{-1}$, $1 < q < {\rm e}^{-\beta\mu}$), it is
possible to decrease the critical temperature $T_c(D)_{qp}$ 
and to get Bose-Einstein condensation 
temperatures in accordance with experiment. However, as a drawback, the model
M$_1$ depends on two parameters $\nu_0^{\prime}$ and $q$. The parameter
$q$ brings a translation factor $ {\beta}^{-1} \ln q $ for the chemical 
potential since  Eq.~(\ref{eq:mu-q})  gives
\begin{equation}
\mu  =  \mu_1 - \frac{\ln q}{\beta}  =  \mu_1 - k_B T \ln q
\label{eq:transl}
\end{equation}
that clearly shows that the importance of the $q$-deformation increases with
the temperature. The parameter  $\nu_0^{\prime}$  is more difficult to
interpret. This parameter is essential, via the factor 
 $ q^{ \frac{2}{D} \nu_0^{\prime} } $   in Eq.~(\ref{eq:(22)}), for decreasing
the critical temperature. In this respect, it might be of importance for
describing some interactions between bosons.  

To close this paper, it is worth noting that alternative choices,
non-linear in $N_k$, are possible for the Hamiltonian 
$H$ in Eq.~(\ref{eq:(3)}). In this direction, 
we may mention, in the case $p=q^{-1}$, the works of 
Refs.~4, 5, 10, 15, and 16. We hope to return 
on these matters in the future, especially on the difficult problem 
of the interpretation of the parameter $ \nu_0^{\prime} $. 

\section*{Acknowledgments}
One of the author (M.R.~K.) would like to thank the Organizing 
Committee of the 4th International School of Theoretical Physics SSPCM96, 
and more specifically Prof.~T.~Lulek, for inviting 
him to deliver this lecture. He is also indebted to Prof.~M.~Bo\.zejko and 
Prof.~Ya.I.~Granovskii for very interesting discussions. 

\section*{Appendix A}
We deal in this appendix with the representation $S(\lambda_0,\nu_0)$ 
of the oscillator algebra $W_q$ with $0 < q <1$. This representation 
follows from 
\begin{eqnarray}
a   \> |n \rangle &=& \left( q^{n}   \lambda_0 
        + q^{-\nu_0} [n  ]_q \right)^{1 \over 2}
    \; |n - 1 \rangle \nonumber\\
a^+ \> |n \rangle &=& \left( q^{n+1} \lambda_0 
        + q^{-\nu_0} [n+1]_q \right)^{1 \over 2}
    \; |n + 1 \rangle \nonumber\\ 
N   \> |n \rangle &=& (\nu_0 + n) \; |n \rangle 
\label{eq:repnonFock} 
\end{eqnarray} 
where $n \in {\bf Z}$ and the real parameters $q$, 
$\lambda_0$ and $\nu_0$ satisfy the constraint equation 
\begin{equation}
q^{-\nu_0} \frac{q}{1-q} < \lambda_0 < q^{-\nu_0} \frac{1}{1-q} 
\label{eq:constrain} 
\end{equation}
A characteristic of this representation is that the spectrum 
of the operator 
$N$ is not bounded from below. It is thus hardly feasible to connect 
$N$ to the usual number operator. Another important characteristic concerns the
absence of a limit of  $S(\lambda_0,\nu_0)$  when $q \to 1^{-}$. In fact, 
the
generic representation $S(\lambda_0,\nu_0)$ disappears at $q = 1$ 
since Eq.~(\ref{eq:constrain}) becomes meaningless for $q \to 1^{-}$. The 
fact that representation $S(\lambda_0,\nu_0)$ has no limit at $q=1$ is probably
the reason why this representation has never been used (to the best of our
knowledge) in physical applications. 

\section*{Appendix B}

We discuss here the consequence of a $qp$-deformation 
of the correlation function $g^{(2)}$ of order two associated to 
the radiation field. In the nondeformed case, we know that $g^{(2)}$ takes 
two values, viz.~$g^{(2)} = 1$ 
for a coherent monomode radiation and $g^{(2)} = 2$
for a chaotic  monomode radiation. An interesting question arises: 
Is it possible to
interpolate between the latter two values by replacing the ordinary boson
operators by $qp$-deformed boson operators? In this appendix, we limit
ourselves to the representation $R(0)$ of the oscillator algebra $W_{qpq}$.

Our basic hypothesis is to describe the radiation field 
by an assembly of photons ($\mu = 0$) with the Hamiltonian
 \begin{equation}
h := \sum_k h_k, \qquad h_k := \hbar \omega_k \nu_k 
 \label{eq:(25)}
 \end{equation}
The corresponding Bose statistical distribution for the $k$ mode is then
given by (\ref{eq:(10)}) 
where $\eta$ is replaced by $\xi := \beta \hbar \omega_k$. 
In the case of $qp$-bosons, we adopt the definition 
 \begin{equation}
g^{(2)} := {\langle a^+a^+aa \rangle \over \langle a^+ a \rangle^2} 
 \label{eq:(26)}
 \end{equation}
where $a$ and $a^+$ stand for the annihilation 
and creation operators for the $k$ mode. 
The abbreviation $\langle  X \rangle$ in 
(\ref{eq:(26)}) denotes the 
mean statistical value $Z^{-1} {\rm tr} \left( {\rm e}^{-\beta h} X \right)$ 
for an operator $X$. Equation (\ref{eq:(26)}) can be developed as
 \begin{equation}
g^{(2)} = 
p^{-1} 
{\langle (a^+ a)^2 \rangle \over \langle a^+ a \rangle^2} -
(qp)^{-1} 
{\langle q^N a^+a  \rangle \over \langle a^+ a \rangle^2} 
 \label{eq:(27)}
 \end{equation}
Of course, $\langle a^+ a \rangle = f_k$ 
as given by (\ref{eq:(10)}) with $\eta \equiv \xi$. 
In addition, the other average values in (\ref{eq:(27)}) can be 
calculated to be
 \begin{equation}
\langle (a^+ a)^2 \rangle =
{({\rm e}^\xi -   1) ({\rm e}^\xi + qp) \over
 ({\rm e}^\xi - q^2) ({\rm e}^\xi - qp) 
 ({\rm e}^\xi - p^2)}
 \label{eq:(28)}
 \end{equation}
and
 \begin{equation}
\langle q^N a^+ a \rangle =
q { {\rm e}^\xi - 1   \over
   ({\rm e}^\xi - q^2) 
   ({\rm e}^\xi - qp )} 
 \label{eq:(29)}
 \end{equation}
Finally, we obtain 
 \begin{equation}
g^{(2)} = (q + p) 
{1 \over {\rm e}^\xi - 1   } 
       {({\rm e}^\xi - q  )^2 
        ({\rm e}^\xi - p  )^2 \over
        ({\rm e}^\xi - q^2) 
        ({\rm e}^\xi - qp ) 
        ({\rm e}^\xi - p^2)} 
 \label{eq:(30)}
 \end{equation}
with convergence conditions that parallel the ones for $f_k$ 
(with $ - 2 \beta \mu $ replaced by $ \xi $). 
[For instance, when $(q,p) \in {\bf R}^+ \times {\bf R}^+$, we must have 
$0 < q < {\rm e}^{\xi \over 2}$ and 
$0 < p < {\rm e}^{\xi \over 2}$.] 
The $qp$-deformed factor $g^{(2)}$ depends on the parameters $q$ 
and $p$ in a symmetrical manner ($q \leftrightarrow p$ symmetry). 
It also presents a dependence on the energy $\hbar \omega_k$ 
of the $k$ mode and on the temperature $T$. 
In the limiting case where $q \to 1$ and $p \to 1$, 
we get $g^{(2)} = 2$ that turns out to be the value 
for the choatic monomode radiation of the black body. 
Let us now examine the cases of low temperatures and high energies. 
In these cases, 
${\rm e}^\xi$ is the dominating term in each of the differences 
occuring in Eq.~(\ref{eq:(30)}). Therefore, we have
 \begin{equation}
g^{(2)} \sim q + p 
 \label{eq:(31)}
 \end{equation}
at low temperature or high energy. 
Equation (\ref{eq:(30)}) shows that we can interpolate 
in a continuous way from  $g^{(2)} = 1$ (coherent phase) to 
                          $g^{(2)} = 2$ (chaotic  phase).
It is thus possible to reach the value $g^{(2)} = 1$
without employing coherent states.  
It should be observed that we can even obtain either 
 $g^{(2)} > 2$ or 
 $g^{(2)} < 1$ from Eq.~(\ref{eq:(30)}). The situation 
where $g^{(2)} < 1$ may be interesting for describing 
antibunching effects of the light field.

\end{document}